\documentstyle[prb,aps,amstex,amssymb,amsfonts,twocolumn,floats,exscale]{revtex}

\begin{document}
\title{The fraction of condensed counterions around a charged rod: \\ 
Comparison of Poisson-Boltzmann theory and computer simulations}
\author{Markus Deserno$^1$, Christian Holm$^1$ and Sylvio May$^2$}
\address{$^1$ Max-Planck-Institut f{\"u}r Polymerforschung, Ackermannweg 10, 
  55128 Mainz, Germany}
\address{$^2$ Institut f{\"u}r Biochemie und Biophysik,
  Friedrich-Schiller-Universit{\"a}t Jena, Philosophenweg 12, 07743 Jena, 
Germany} 
\date{\today}
\maketitle


\renewcommand{\thefootnote}{\fnsymbol{footnote}}


\newcommand{\CC}{{\mathbb{C}}}

\newcommand{\infd}{{\operatorname{d}}}
\newcommand{\rB}{{\operatorname{B}}}
\newcommand{\rb}{{\operatorname{b}}}
\newcommand{\rD}{{\operatorname{D}}}
\newcommand{\rM}{{\operatorname{M}}}
\newcommand{\rT}{{\operatorname{T}}}


\hspace{2cm}\begin{abstract}
  \begin{centering} \parbox{16cm}{ \vspace{-1cm}\hfill\parbox{14.2cm}{
        We investigate the phenomenon of counterion condensation in a solution
        of highly charged rigid polyelectrolytes within the cell model.  A
        method is proposed which -- based on the charge distribution function
        -- identifies both the fraction of condensed ions and the radial
        extension of the condensed layer. Within salt-free Poisson-Boltzmann
        (PB) theory it reproduces the well known fraction $1-1/\xi$ of
        condensed ions for a Manning parameter $\xi>1$.  Furthermore, it
        predicts a weak salt dependence of this fraction and a breakdown of
        the concept of counterion condensation in the high salt limit.  We
        complement our theoretical investigations with molecular dynamics
        simulations of a cell-like model, which constantly yield a stronger
        condensation than predicted by PB theory. While the agreement between
        theory and simulation is excellent in the monovalent, weakly charged
        case, it deteriorates with increasing electrostatic interaction
        strength and, in particular, increasing valence. For instance, at a
        high concentration of divalent salt and large $\xi$ our computer
        simulations predict charge oscillations, which mean-field theory is
        unable to reproduce.  }}
  \end{centering}
\end{abstract}



\narrowtext


\section{Introduction}

Strongly charged linear polyelectrolytes use their counterions to reduce their
line charge density.\cite{manning96a}  This phenomenon has led to the concept
of {\em counterion condensation},\cite{manning69a,oosawa70a} and although it
was introduced a long time ago, varying viewpoints about this subject persist
in the literature.\cite{manning96a,Beispiele}  Here we will investigate its
appearance within the commonly used {\em cell
  model}\/\cite{lifson53a,katchalsky71a}: an infinitely long charged rod
enclosed in a cylindrical cell together with its counterions --- with and
without added salt.

In the salt-free case this model can be solved analytically within nonlinear
Poisson-Boltzmann (PB) theory,\cite{alfrey51a,fuoss51a} which thus affords a
particularly clear view on Manning-Oosawa counterion condensation, as has been
demonstrated by Zimm and Le Bret.\cite{zimm83a,lebret84a}  Going beyond
salt-free PB theory, many questions arise: How closely condensed and tightly
bound is the ``condensed layer''?  What distinguishes condensed from
uncondensed counterions?  When is the {\em mean-field} level PB theory a good
approximation for real systems? And how does the presence of salt affect the
condensation phenomenon?

Recently Manning proposed the idea that there exists a clear distinction
between {\em (i)} a condensed layer and {\em (ii)} a distant, more diffuse
``Debye-H\"uckel'' cloud.\cite{manning98a} In the integrated radial counterion
distribution function this is supposed to be detectable as an inflection
point, which separates the two regions. Here we argue that this is not quite
in accord with PB theory without added salt: There is an inflection point in
the distribution function, but it is not related to the condensation
phenomenon.  If however the distribution function is plotted against {\em
  logarithmical} radial distance, an inflection point appears which exactly
divides the counterions into condensed and uncondensed ones, as previously
pointed out by Belloni.\cite{belloni84a}  Since this feature has largely gone
unnoticed in the study of polyelectrolytes, we found it worthwhile to present
its derivation within PB theory in a short and explicit form and also to point
out its {\em practical} usability. To this end, we present computer
simulations, compare them to PB theory, and demonstrate that this criterion
can indeed be extended to quantify counterion condensation even beyond the
scope of PB theory.

\begin{figure}[t]
  \vspace{6.6cm}
\end{figure}

In the case of added salt the analytical treatment of the cylindrical PB
equation is much more involved,\cite{Rama85,Tracy97} so any potential
inflection point criterion is more difficult to analyze.  Still, the PB
equation can be solved numerically, suggesting that the presence of monovalent
salt decreases the {\em extension} of the condensed layer, but leaves the {\em
  amount} of condensed counterions largely unaffected.  We show that this
behavior is well reproduced in computer simulations for monovalent salt.  We
also present a simple criterion for determining the salt concentration above
which the concept of Manning condensation is no longer meaningful.  In the
regime of strong electrostatics, high valence and much added salt the PB
predictions deviate qualitatively from simulational results. In particular,
the simulation shows a pronounced overcharging and charge oscillations, which
are absent on the {\em mean-field} PB level.

This paper is structured as follows: In sections~\ref{sec_PBwithout} and
\ref{sec_coucond} we recapitulate the main ingredients of the PB solution for
the salt-free cell model and illustrate the connection of Manning condensation
with the inflection point criterion mentioned above. This is followed -- in
section~\ref{sec_compwithout} -- by a comparison of the salt-free PB results
with computer simulations.  In sections~\ref{sec_coucondwith} and
\ref{sec_PBwith} we discuss the concept of counterion condensation in the
presence of salt and to this end derive a PB equation for the ensemble of
constant number of salt molecules. Its results are compared with simulations
in section~\ref{sec_compwith}. Details of our simulation method can be found
in the appendix.


\section{PB-theory for a charged rod without added salt}\label{sec_PBwithout}

Consider an infinitely long cylinder of radius $r_0$ and line charge density
$\lambda>0$, which is coaxially enclosed in a cylindrical cell of radius $R$.
Global charge neutrality of the system is ensured by adding an appropriate
amount of oppositely charged (monovalent) counterions.

Within PB theory these counterions are replaced by a cylindrically symmetric
counterion {\em density} $n(r)$ ($r$ is the radial coordinate) which gives
rise to an electrostatic potential $\Phi(r)$ satisfying the Poisson equation
\begin{equation}\label{Poisson}
\left(\frac{\infd^2}{\infd r^2}+\frac{1}{r}\frac{\infd}{\infd r}\right)\Phi(r)
= \frac{e}{\epsilon} \, n(r)
\end{equation}
with $\epsilon$ being the dielectric constant outside the
cylinder\cite{cylinderepsilon} and $e$ the (positive) unit of charge.
Conversely, this potential is supposed to influence the counterion density via
the Boltzmann factor:
\begin{equation}\label{Boltzmann}
n(r) = n(R)\,\exp\big\{\beta e \Phi(r)\big\}
\end{equation}
with the inverse temperature $\beta = 1/k_\rB T$ and $k_\rB$ being Boltzmann's
constant. Thus, the chosen zero of the potential is $\Phi(R)=0$.

In the following it is advantageous to change variables: The Bjerrum
length $\ell_\rB = \beta e^2/4\pi\epsilon$ provides a convenient
scale for quantifying electrostatic interactions (it is the distance,
at which the Coulomb energy of two elementary charges equals $k_\rB
T$), while the dimensionless Manning parameter $\xi = \lambda
\ell_\rB / e$ measures the line charge density of the rod (it is equal
to the number of elementary charges per Bjerrum length).  We shall
only be interested in the strongly charged case $\xi>1$.  Insertion of
eq~\ref{Boltzmann} into eq~\ref{Poisson} results in the
nonlinear PB equation which in terms of the reduced (dimensionless)
electrostatic potential $y(r) = \beta e \Phi(r)$ and a screening
length $1/\kappa>0$ with $\kappa^2 = 4\pi\ell_\rB n(R)$ reads
\begin{equation}\label{Poisson_Boltzmann}
y'' + \frac{y'}{r} = \kappa^2 e^y
\end{equation}
The appropriate boundary conditions for solving the PB equation arise from
applying Gau{\ss}' law at $r_0$ and $R$:
\begin{equation}\label{boundary_E_field}
y'(r_0) = -2\xi/r_0 \qquad,\qquad y'(R) = 0
\end{equation}

\noindent
Eq~\ref{Poisson_Boltzmann} has an analytical solution which can be written
in the following way:
\begin{equation}\label{PB_potential}
y(r) = -2\,\ln\left\{\frac{\kappa \, r}{\gamma \sqrt{2}}\cos
  \Big(\gamma\,\ln\frac{r}{R_\rM}\Big)\right\}.
\end{equation}

\noindent
The boundary conditions (\ref{boundary_E_field}) yield two coupled,
transcendental equations for the two integration constants $\gamma$ and
$R_\rM$:
\begin{eqnarray}
\gamma\,\ln\frac{r_0}{R_\rM} & = & \arctan\frac{1-\xi}{\gamma}
\label{r0_RM} \\
\gamma\,\ln\frac{R}{R_\rM} & = & \arctan\frac{1}{\gamma} \label{R_RM}
\end{eqnarray}
Subtracting (\ref{r0_RM}) from (\ref{R_RM}) eliminates $R_\rM$ and provides an
equation from which $\gamma$ can be obtained {\em numerically}. The second
integration constant $R_\rM$, which we will refer to as the {\em Manning
  radius}, is then given by either of these equations.  Note also that
$\kappa$ and $\gamma$ are connected via $\kappa^2R^2 = 2 \, (1+\gamma^2)$,
thus ensuring the chosen normalization of the potential.

The Manning radius $R_\rM$ depends monotonically on $\xi$ and for $\xi > 1$
one finds $R_\rM > r_0$. As discussed in the next section this is the regime
in which counterion condensation occurs. If $\xi = 1$ then $R_\rM = r_0$,
i.e., the Manning radius is located at the surface of the rod.  A further
decrease in $\xi$ shifts $R_\rM$ inside the cylinder and for
$\xi=\ln(R/r_0)/(1+\ln(R/r_0))$ both the Manning radius and $\gamma$ vanish.
Even smaller values of $\xi$ render the integration constant $\gamma$ complex.
Still, the solution (\ref{PB_potential}) can be extended by analytic
continuation over $\CC$.

Using eqs \ref{Boltzmann}, \ref{PB_potential} and \ref{r0_RM} the total
charge per unit length, $Q(r)$, found within a cylinder of radius
$r\in[r_0;R]$ can be determined by integration:
\begin{eqnarray}
Q(r)/\lambda & = & 1 - \frac{1}{\lambda} \int_{r_0}^r \infd \bar{r} \, 
2\pi \bar{r} \, 
e \, n(\bar{r}) \nonumber \\
{} & = & 1 - \Big(1 - \frac{1}{\xi}\Big) -
\frac{\gamma}{\xi}\tan\Big(\gamma\,\ln\frac{r}{R_\rM}\Big) 
\label{integrated_charge}
\end{eqnarray}
Since $n(r)>0$, $Q(r)$ decreases monotonically from $Q(r_0)=\lambda$ to
$Q(R)=0$. The latter follows from eq~\ref{R_RM} and is a consequence of
global charge neutrality. It is instructive to use the quantity
\begin{equation}
P(r) = 1-Q(r)/\lambda,\label{probdistri}
\end{equation}
which is the probability of finding a mobile ion within the distance $r$.  In
other words, it is the {\em fraction of counterions} found within a cylinder
of radius $r$.  In particular, at $r=R_\rM$ the last term in $Q$, as given in
eq~\ref{integrated_charge}, vanishes, giving a fraction $1-1/\xi$ of ions
within $R_\rM$. It can easily be verified that generalizing
eqs~\ref{boundary_E_field}, \ref{r0_RM} and \ref{integrated_charge} for
counterions with valence $v$ reduces to replacing $\xi\rightarrow\xi v$.
Within PB-theory changing valence or electrostatic interaction strength
affects the charge distribution function in the same way.


\section{Counterion condensation: Definition and identification}\label{sec_coucond}

\noindent
For $\xi > 1$ eqs~\ref{r0_RM} and \ref{R_RM} imply the inequalities
\begin{equation}
\frac{\pi}{\ln\frac{R}{r_0}} 
\; \ge \; \gamma \; \ge \; 
\frac{\pi}{\ln\frac{R}{r_0}+\frac{\xi}{\xi-1}}\label{gamma_asymptotic}.
\end{equation}
Since the two bounds become equal in the limit $R\rightarrow\infty$, they
provide an asymptotic solution for $\gamma$ and give rise to various limiting
laws, which illuminate the behavior of the solution in the dilute limit. In
particular, the reduced potential for $\xi>1$ becomes:\cite{Joanny}
\begin{equation}
  \hspace*{-4ex}
  y(r)-y(r_0) = -2\,\ln\frac{r}{r_0} - 
  2\,\ln\Big\{1+(\xi\!-\!1)\ln\frac{r}{r_0}\Big\},
\end{equation}
which is (up to a logarithmic correction) identical to the potential of a rod
with charge density of $e/\ell_\rB$, i.e., Manning parameter $\xi=1$. This can
be attributed to a condensation of counterions onto the rod, which renormalize
the line charge density. Indeed, the contact density $n(r_0)$ converges to the
nonzero value
\begin{equation}\label{contact_density}
\lim_{R\rightarrow\infty} n(r_0) = \frac{\lambda}{\pi
  r_0^2 e}\frac{(\xi-1)^2}{2\xi},
\end{equation}
suggesting the existence of a close layer which cannot be diluted away.

In order to establish an effective Manning parameter of $1$, a fraction $f_\xi
= 1-1/\xi$ of all counterions would have to condense onto the rod. In fact,
$f_\xi$ is a critical threshold in the following sense: For $0<\alpha<1$,
$\xi>1$ and the radius $r_\alpha$ defined as $r_\alpha = r_0 \,
\exp\left\{\alpha/(\xi-1)(1-\alpha)\right\}$ one can show by using
eqs~\ref{r0_RM}--\ref{probdistri} that
\begin{equation}\label{Manning_criticality}
\lim_{R\rightarrow\infty} P(r_\alpha) = \alpha\,f_\xi.
\end{equation}
Hence, in the limit of infinite dilution a fraction $\alpha$ (arbitrarily
close to $1$) of the fraction $f_\xi$ stays within a {\em finite} radius
$r_\alpha$. It has thus been common practice to call $f_\xi$ the {\em fraction
  of condensed counterions} or {\em Manning fraction}, although
$\lim_{\alpha\rightarrow 1} r_\alpha=\infty$. Actually, $R_\rM$ diverges like
$\sqrt{R}$, which follows directly from either of the asymptotic boundaries
\ref{gamma_asymptotic} for $\gamma$.\cite{lebret84a,Gueron80}

\vspace{0.3cm}

Investigating counterion condensation by means of computer simulations
requires a criterion which identifies condensed ions. Here we show that the
{\em functional form} of the counterion distribution function suggests a
simple rule for recognizing the Manning layer.

If the function $P$ is known, the condensed counterion fraction can be
characterized in the following ``geometric'' way:
Eq~\ref{integrated_charge} shows that $P$ viewed as a function of $\ln(r)$
is merely a {\em shifted tangent-function} with its center of symmetry at
$\{\ln(R_\rM);f_\xi\}$. Since $\tan''(0) = 0$, the Manning radius and Manning
fraction can be found by plotting $P$ as a function of $\ln(r)$ and {\em
  localizing the point of inflection}.

This property of $P$, derived within the framework of PB-theory, can in turn
be used to {\em define} the condensed fraction.  It provides a suitable way to
quantify coun\-ter\-ion condensation beyond the scope of PB-theory, and it is
exact in the PB-limit without added salt, by construction.

Our counterion condensation criterion can be reformulated in terms of the
counterion density $n(r)$: If $P$ has a point of inflection as a function of
$\ln r$, $\infd P/\infd\,\ln(r)$ must have a stationary point there. Using
eqs~\ref{integrated_charge} and \ref{probdistri} it follows that $r^2\,n(r)$
must have a stationary point, which in the simple salt-free case is actually a
minimum. (In fact, for our simulated data we localized the point of inflection
by fitting a (2,2) Pad\'{e} approximant in $\ln r$ to $r^2\,n(r)$ in the
vicinity of its minimum.)

It is appropriate to mention briefly here three other methods which have been
used to measure counterion condensation and point out their shortcomings.  The
notion of a condensed layer closely surrounding the rod suggests determining
the condensed fraction by simply counting the ions within a certain (small)
distance of the rod, say, a few diameters or one screening length
$1/\kappa$.\cite{Mandel92} This amounts to making a prior assumption about the
Manning radius. Such a procedure is not only arbitrary; moreover, the PB
Manning radius depends on the polyelectrolyte density and diverges like
$\sqrt{R}$ in the dilute limit.  If this is not taken into account, the
condensed fraction is either underestimated (for fixed condensation distance)
or overestimated (for a distance proportional to the screening length
$1/\kappa$ of the counterions, which is proportional to $R$).

Conversely, one could assume that the condensed fraction is {\em always} given
by $1-1/\xi$ and thereby obtain the size of the condensed layer, e.g.\ when
salt is added to the system.\cite{Gueron80} Although being exact in the
salt-free PB limit, this criterion excludes by definition the possibility that
any effects beyond the {\em mean-field} level (like correlations) or the
presence of salt also modify the {\em fraction} of condensed counterions.  It
also does not predict a crossover to a high salt regime where all counterions
are condensed solely due to the presence of the salt (see
section~\ref{sec_PBwith}).

\begin{figure}
  \vspace*{-0.5cm}
  \begin{center} 
\setlength{\unitlength}{0.1bp}
\begin{picture}(2339,1728)(0,0)
\special{psfile=fig1 llx=0 lly=0 urx=468 ury=403 rwi=4680}
\put(1344,50){\makebox(0,0){$R/r_0$}}
\put(50,964){%
\special{ps: gsave currentpoint currentpoint translate
270 rotate neg exch neg exch translate}%
\makebox(0,0)[b]{\shortstack{condensed fraction}}%
\special{ps: currentpoint grestore moveto}%
}
\put(2339,200){\makebox(0,0){1000}}
\put(1992,200){\makebox(0,0){300}}
\put(1676,200){\makebox(0,0){100}}
\put(1329,200){\makebox(0,0){30}}
\put(1013,200){\makebox(0,0){10}}
\put(666,200){\makebox(0,0){3}}
\put(350,200){\makebox(0,0){1}}
\put(300,1628){\makebox(0,0)[r]{1.0}}
\put(300,1362){\makebox(0,0)[r]{0.8}}
\put(300,1097){\makebox(0,0)[r]{0.6}}
\put(300,831){\makebox(0,0)[r]{0.4}}
\put(300,566){\makebox(0,0)[r]{0.2}}
\put(300,300){\makebox(0,0)[r]{0.0}}
\end{picture}

 \end{center}
  \caption{Three predictions for the fraction of condensed counterions for a
    cell with $r_0=\sigma$ and $\xi=2$ as a function of cell size $R$. The
    solid line is the inflection point criterion. The dashed line derives the
    distinction between condensed and uncondensed ions from the condition
    $y(r)=1$ while the dotted line takes $y(r_0)-y(r)=1$. The two arrows mark
    the values for the energy based criteria in the limit
    $R\rightarrow\infty$.  Notice that they do not coincide with the Manning
    fraction $1-1/\xi=1/2$.}\label{fig:1}
\end{figure}
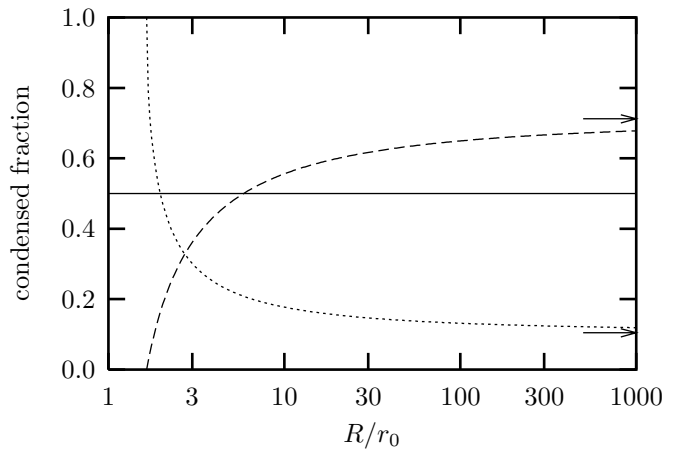

Finally, one could be tempted to bring into play the {\em electrostatic
  binding energy} and regard all ions within a thermal distance $R_\rT$
defined by $y(R_\rT)=1$ as condensed.\cite{Pack93,Lamm94} (Alternatively, one
might require a potential difference of $k_\rB T/e_0$ with respect to the rod
surface, i.e., $y(r_0)-y(R_\rT')=1$.) Within salt-free PB theory, however,
this is not a suitable criterion since the value of the electrostatic
potential at the Manning radius, $\Phi(R_\rM)$, is in no way special (upon
dilution it actually diverges logarithmically with respect to the boundary as
well as the rod surface, as can be derived from
eq~\ref{gamma_asymptotic}).\cite{KTspecial} As an illustration,
figure~\ref{fig:1} compares the inflection point rule and two energy
based criteria with regard to their predictions for the condensed fraction:
All three methods quantify condensation differently, and the energy based
approaches are density dependent -- which on its own is not a problem. The
unsatisfying aspect is rather that the latter do not converge against the
Manning fraction $1-1/\xi$ upon dilution. In fact, using
eq~\ref{gamma_asymptotic} it can e.g.\ be shown that $R_\rT\thicksim c_1 \, R$
in the dilute limit, with $c_1 \approx 0.25726$ being a solution of
$c_1(1-\ln\,c_1)=e^{-1/2}$. Observe that $R_\rT$ scales {\em linearly} with
$R$ and thus {\em faster} than the Manning radius; therefore, the thermal
radius will enclose {\em more} than the Manning fraction in the dilute limit.
In fact, $P(R_\rT)\thicksim 1-c_2/\xi$ with the constant $c_2 =
\ln\,c_1/(\ln\,c_1 - 1)\approx 0.57585$.  Incidentally, this implies the
relative deviation of this approach from the Manning fraction $1-1/\xi$ to
become small at large $\xi$.

Let us thus repeat that the inflection point criterion employed in the present
work has the advantages of $(i)$ not fixing by definition the amount of
condensed counterions ($f_\xi$ {\em and} $R_\rM$ can be determined
independently of each other), $(ii)$ reproducing the salt-free PB limit,
namely $P(R_\rM)=1-1/\xi$, and $(iii)$ quantifying the breakdown of the
coexistence of condensed and uncondensed counterions in the high salt limit,
as will be shown in Sec.~\ref{sec_PBwith}.

In concluding this section we note that the appearance of the logarithm in the
inflection point criterion is related to $\ln(r)$ being the 2D Coulomb
potential, i.e., the {\em Green function} of the cylindrically symmetric
Laplacian. In the corresponding 3D (spherical) problem of charged colloids the
Green function $1/r$ would be the appropriate choice for plotting the radial
coordinate.\cite{belloni84a}


\section{Comparison of PB theory with simulations: No added salt}\label{sec_compwithout}

In this section we supplement the results of salt-free PB theory with computer
simulations of a cell-like model, with particular emphasis on the role of
Manning parameter and valence.  Details of the model, the simulations and our
notation conventions are summarized in the appendix.

Figure~\ref{fig:2} shows the counterion distribution functions,
$P(r)$, for three systems with monovalent counterions, $l_\rB/r_0=1$,
$R/r_0=123.8$ and $\xi\in\{0.96,1.92,2.88\}$, i.e., counterion condensation is
expected to occur for the latter two. As suggested in the previous section the
functions are plotted using a logarithmically scaled $r$-axis. Note that in
all our PB calculations and simulations the distance of closest approach to
the rod was $r_0=\sigma$, where $\sigma$ is the small ion diameter, used in
the simulations (see appendix).

\begin{figure}
  \vspace*{-0.5cm}
  \begin{center} 
\setlength{\unitlength}{0.1bp}
\begin{picture}(2339,1728)(0,0)
\special{psfile=fig2 llx=0 lly=0 urx=468 ury=403 rwi=4680}
\put(1344,50){\makebox(0,0){$r/r_0$}}
\put(50,964){%
\special{ps: gsave currentpoint currentpoint translate
270 rotate neg exch neg exch translate}%
\makebox(0,0)[b]{\shortstack{$P(r)$}}%
\special{ps: currentpoint grestore moveto}%
}
\put(2217,200){\makebox(0,0){100}}
\put(1949,200){\makebox(0,0){50}}
\put(1595,200){\makebox(0,0){20}}
\put(1327,200){\makebox(0,0){10}}
\put(1059,200){\makebox(0,0){5}}
\put(704,200){\makebox(0,0){2}}
\put(436,200){\makebox(0,0){1}}
\put(300,1628){\makebox(0,0)[r]{1.0}}
\put(300,1362){\makebox(0,0)[r]{0.8}}
\put(300,1097){\makebox(0,0)[r]{0.6}}
\put(300,831){\makebox(0,0)[r]{0.4}}
\put(300,566){\makebox(0,0)[r]{0.2}}
\put(300,300){\makebox(0,0)[r]{0.0}}
\end{picture}

 \end{center}
  \caption{Simulated counterion distributions (solid lines) and PB-results
    (dotted lines) for a monovalent system with $r_0=1\,\sigma$, $R/r_0
    \approx 123.8$ and (from bottom to top) $\xi\in\{0.96,1.92,2.88\}$. The
    $\uparrow$-arrows mark the inflection points in the PB-distribution while
    the $\downarrow$-arrows mark those points in the MD distributions. Note
    the logarithmically scaled $r$-axis in the present and the following
    figures.}\label{fig:2}
\end{figure}
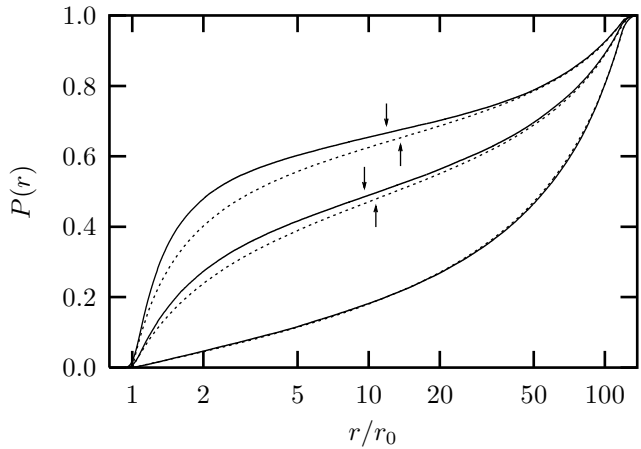

For the system with $\xi=0.96$ in figure~\ref{fig:2} the agreement
between simulation and PB theory is excellent -- deviations are almost within
the linewidth of the plotted curves. For the other two cases the agreement
fails quantitatively, but {\em not} qualitatively: The {\em shape} of the
distribution function remains largely unchanged.  The curves for $P(r)$ lie
above the PB result for all $r$, indicating a stronger condensation than
predicted by {\em mean-field} theory. Importantly, for $\xi>1$ the simulated
curves display a point of inflection.  As described in the previous section
this can be used to define a Manning radius and a condensation fraction, which
permits one to quantify by {\em how much} the condensation is stronger. This
is summarized for a range of densities in table~1, where it can
be seen that deviations towards higher condensation are stronger for dense
systems and relax towards the PB prediction upon dilution.

\begin{table}[b]
  \begin{tabular}{c|ccccccc}
    $R/r_0$    & 2.06 & 3.87 & 7.74 & 15.5 & 31.0 & 62.0 & 124  \\ \hline
    $\xi=1.92$ & 0.56 & 0.57 & 0.56 & 0.54 & 0.53 & 0.51 & 0.49 \\
    $\xi=2.88$ & 0.78 & 0.76 & 0.73 & 0.70 & 0.69 & 0.67 & 0.67 \\
  \end{tabular}
  \vspace*{3ex}
  \caption{Measured condensation fraction $f_\xi$ for various monovalent systems
    with $r_0=\sigma$, which differ in cell size $R$ and thus polyelectrolyte
    density. Within PB theory $f_\xi=1-1/\xi$, giving $f_{1.92} \approx 0.479$
  and $f_{2.88} \approx 0.653$ -- independently of $R$. Note that the
  counterion distribution functions for $R/r_0=124$ are shown in
  figure~\ref{fig:2}.}\label{cond_table}
\end{table}

Note that in figure~\ref{fig:2} the measured fraction $f_\xi$ is {\em
  larger} than the PB-prediction, while the corresponding Manning radius is
{\em smaller}; but since within PB-theory $R_\rM$ increases monotonically with
$\xi$, the measured curves cannot be modelled by a PB distribution with a
somewhat larger {\em effective} $\xi$.

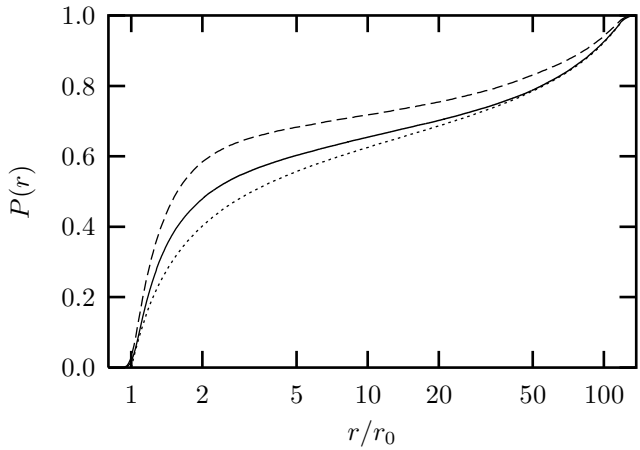
\begin{figure}
  \vspace*{-0.5cm}
  \begin{center} 
\setlength{\unitlength}{0.1bp}
\begin{picture}(2339,1728)(0,0)
\special{psfile=fig3 llx=0 lly=0 urx=468 ury=403 rwi=4680}
\put(1344,50){\makebox(0,0){$r/r_0$}}
\put(50,964){%
\special{ps: gsave currentpoint currentpoint translate
270 rotate neg exch neg exch translate}%
\makebox(0,0)[b]{\shortstack{$P(r)$}}%
\special{ps: currentpoint grestore moveto}%
}
\put(2217,200){\makebox(0,0){100}}
\put(1949,200){\makebox(0,0){50}}
\put(1595,200){\makebox(0,0){20}}
\put(1327,200){\makebox(0,0){10}}
\put(1059,200){\makebox(0,0){5}}
\put(704,200){\makebox(0,0){2}}
\put(436,200){\makebox(0,0){1}}
\put(300,1628){\makebox(0,0)[r]{1.0}}
\put(300,1362){\makebox(0,0)[r]{0.8}}
\put(300,1097){\makebox(0,0)[r]{0.6}}
\put(300,831){\makebox(0,0)[r]{0.4}}
\put(300,566){\makebox(0,0)[r]{0.2}}
\put(300,300){\makebox(0,0)[r]{0.0}}
\end{picture}

 \end{center}
  \caption{Comparison of Manning parameter and valence. The simulated system
    from figure~\ref{fig:2} with $\xi=2.88$ (solid curve) and its PB-solution
    (dotted curve) are contrasted with a simulation where the Manning
    parameter is three times smaller but the counterions are trivalent (dashed
    curve).}\label{fig:3}
\end{figure}

As already mentioned, within PB theory the shape of the integrated
distribution function, $P(r)$, depends on the Manning parameter $\xi$ and the
counterion valence $v$ only via the pro\-duct $\xi v$. We now show that this
is a property seen on the {\em mean-field} level only.  In addition to the PB
result and the simulation of the system with $\xi=2.88$ already shown in
figure~\ref{fig:2}, the simulated $P(r)$ curve for a system with {\em
  trivalent} counterions and $\xi=2.88/3=0.96$ is displayed in
figure~\ref{fig:3}. The corresponding $P(r)$ indicates an even stronger
condensation than the simulation for $\xi=2.88$ and monovalent counterions.

There are two principle ways in which PB theory can fail: {\em (i)} the
neglect of excluded volume interactions (``point-like'' ions) and {\em (ii)}
missing correlations. For the first point there is a simple self-consistency
check: The counterion density is highest at the surface of the inner rod and
e.g.\ in the limit $R\rightarrow\infty$ given by eq~\ref{contact_density}, but
counterions with a finite size might not be able to give rise to a density as
large as that.  This consequently limits the range of applicability of the PB
approximation towards not too large ions, not too small cylinders and not too
strong electrostatics. In our simulations we are below that limit.  Excluded
volume interactions -- if present -- reduce the contact
density\cite{Borukhov97,Lue99} and hence could only lower the integrated
distribution function $P(r)$; this however is not found in the simulations.
The observed stronger condensation must therefore be attributed to
correlations neglected in the PB approach.  Since an increase in density goes
along with an increase of correlations, this explanation seems to be
intuitively correct and it is also supported by quasi-analytical theories,
which go beyond the {\em mean-field} level.\cite{tovar85a,Das95Das97}


\section{Counterion condensation in the presence of salt}\label{sec_coucondwith}

The central question to be discussed in this section is the following: Can the
concept of counterion condensation be extended to the case of added salt?

Note first of all that the salt corresponds to a new degree of freedom which
comes along with its own length scale, namely, a Debye length
$\ell_\rD=(8\pi\ell_\rB v^2 n)^{-1/2}$, where $v$ is the valence of the (for
simplicity) symmetric salt and $n$ is its density. It is of central importance
how this new length relates to the characteristic length $R_\rM$ of the
condensation structure: If $\ell_\rD$ is large compared to $R_\rM$, the
pure-counterion $n(r)$ structure is preserved; if $\ell_\rD$ is smaller, it
dictates the shape of the charge distribution function and the condensation
structure is no longer present.

Since counterion condensation becomes apparent in the behavior of the charge
distribution function for $R\rightarrow\infty$, one should also investigate
this limit in the presence of salt. This however is crucially dependent on the
chosen {\em ensemble}, i.e., whether the limit is performed at constant number
$N$ of salt molecules or at constant chemical potential $\mu$.

In the constant $N$ case $\ell_\rD$ is proportional to $R$ and thus diverges
faster than the Manning radius $R_\rM$, which only scales like $\sqrt{R}$.
For sufficiently large $R$ the condensation structure will therefore be
visible and the condensation criterion will be the same as in the salt-free
case.  Conversely, for sufficiently high density or number of salt molecules,
$\ell_\rD$ will be {\em smaller} than the Manning radius, thus modifying the
condensation structure. Since the latter is a new mechanism for compensating
the rod charge, it is no longer sensible to use the concept of Manning
condensation in this limit. It remains the task of clarifying the crossover
from {\em counterion condensation} to {\em screening}, which is subject of the
following section.

This line of reasoning needs a little modification if the added salt has a
higher valence than the counterions, since then it will preferably be the salt
ions which will condense onto the rod. Two cases have to be distinguished:
\begin{enumerate}
\item Already a fraction of the negative salt ions of highest valence could
  completely neutralize the rod. If these ions are taken to be the ``true''
  counterions and all the rest (including the ``original'' counterions) is
  denoted as ``salt'', one can expect a Manning limiting behavior typical for
  the highvalent new counterions.
\item There is not enough salt to completely neutralize the rod with the
  negative salt ions. This is just as complicated as the salt-free case with
  different species of counterions and will not be pursued further in this
  paper.
\end{enumerate}

\noindent
Quite differently, in the constant $\mu$ case the Debye length of the salt
will remain {\em finite} in the limit $R\rightarrow\infty$ and consequently
smaller than the diverging Manning radius. The condensation structure will
always be wiped out in the infinite dilution limit and it is not possible to
produce a condensation criterion along the lines of the salt-free case. We
therefore prefer to work in the constant $N$ ensemble.

From an experimentalists point of view, keeping $N$ or $\mu$ constant in the
limit $R\rightarrow\infty$ corresponds to two completely different procedures:
In the first case the polyelectrolyte solution is diluted by the addition of
pure water. In the second case the dilution is done with a salt solution of
the same ionic strength as the one in which the polyelectrolyte originally has
been dissolved. One therefore cannot expect these two cases to become
equivalent in the thermodynamic limit.

\vspace{1ex}

In large systems it is essentially irrelevant whether a certain salt
concentration is achived by choosing a certain {\em number} of salt ions or a
corresponding {\em chemical potential} for them. Yet, in all typical systems
accessible to computer simulations the number of ions is still rather small,
so that the chosen ensemble matters. Since eventually a comparison between
simulation and theory is headed for and since the most straightforward
ensemble for simulations is the one which conserves particle number, the
following section derives a PB equation in the presence of salt for this case.


\section{Poisson-Boltzmann equation for constant number of salt molecules}\label{sec_PBwith}

Assume that in addition to the monovalent counterions of the positively
charged rod the cell contains $K$ different $v:v$ salts of concentrations
$\bar{n}_v$ with $v=1 \dots K$. The overall concentration of negative
monovalent ions is thus $\bar{n}_1+m$ with $m=\lambda/e \pi R^2$ where $m$ is
the contribution due to the counterions of the rod.

The free energy $F=U-TS$ accounts for the internal electrostatic energy 
$U$ and the translational entropy $S$ of the mobile ions
in solution.  It can be written in terms of the electrostatic
potential $\Phi$ and the local ion concentrations $n_v$ and $n_{-v}$ of
positive and negative ions of valence $v$, respectively.  
Within {\em mean-field} theory, $F$ is given by
\begin{equation}
F=\int \limits_V \infd^3 r\; \bigg[ \frac{\epsilon}{2} \big(\nabla
  \Phi\big)^2+k_\rB T \sum \limits_{v=-K \atop v \neq 0}^{K} 
n_v \ln \frac{n_v}{\bar{n}_v}
\bigg]
\end{equation}
where $\bar{n}_{-v}=\bar{n}_v$ (for $v=2 \dots K$) and
$\bar{n}_{-1}=\bar{n}_1+m$ denote the concentrations of the negatively 
charged mobile ions.

As discussed in the above section, we are interested in the constant $N$
ensemble, i.e., the case that for each ionic species the number of ions within
the cell of volume $V$ is conserved. The local equilibrium concentrations
$n_v$ have thus to be derived under the constraints
\begin{equation}
\langle n_v \rangle \equiv \frac{1}{V} \int \limits_V \infd^3 r \; n_v = 
\bar{n}_v
\end{equation}
The usual variation of $F$ results then in the Boltzmann distributions
for the local concentrations
\begin{equation}
n_v=\bar{n}_v e^{-v y-\mu_v}
\end{equation}
where the chemical potentials $\mu_v=\ln \langle e^{-v y} \rangle$ ensure
particle conservation.

Again we consider the rod sufficiently long that we can neglect end effects.
Then, the electrostatic potential $\Phi$ and the local ion concentrations
$n_v$ depend only on the radial distance $r$ to the rod axis.

Insertion of the local concentrations $n_v$ into the cylindrically symmetric
Poisson equation $\epsilon\,(\Phi''+\Phi'/r)=-\sum_{v} v n_v$ leads to the
Poisson-Boltzmann equation
\begin{equation} \label{pbe}
y''+\frac{y'}{r}=-4 \pi \ell_\rB \sum_{v=-K}^{K} v \bar{n}_v
\frac{e^{-v y}}{\langle e^{-v y} \rangle}
\end{equation}
This equation has to be solved subject to the boundary conditions
(\ref{boundary_E_field}), i.e.\ the same as for the salt-free case.

Numerical solutions of eq~\ref{pbe} can be found employing a Newton-Raphson
iteration scheme in which the chemical potentials $\mu_v=\ln \langle e^{-v y}
\rangle$ are updated after each iteration step.  Once a solution $y(r)$ is
found, the {\em integrated charge distribution function} of the mobile ions
\begin{equation} \label{it4}
P(r)=\frac{e}{\lambda} \int \limits_{r_0}^r \infd r' \: 2 \pi r' \sum
\limits_{v=-K}^K  v n_v(r)
\end{equation}
can simply be calculated by $P(r)=1+r y'(r)/2 \xi$ which follows from
inserting the Poisson equation into eq~\ref{it4} and carrying out the
integration with consideration of the boundary conditions
(\ref{boundary_E_field}).  Since $P(r)$ is a measure of the fraction of the
overall electrolyte charge found within a cylinder of radius $r$, we must have
$P(r_0)=0$ and $P(R)=1$.  Note that eq~\ref{it4} is a natural
generalization of the distribution function from eq~\ref{probdistri}, but
its interpretation as an integrated probability distribution (or fraction of
counterions) is only valid in the salt-free case.

To investigate the condensation criterion in the presence of monovalent salt
we have calculated the {\em mean-field} potential $y(r)$ solving the PB
equation for a system characterized by $r_0=\sigma$, $R/r_0=61.9$,
$\lambda/e=0.96/r_0$, $\xi=2.1$ and a variable number of salt molecules. To
facilitate the comparison with computer simulations in the next section we
write $N$ for the number of monovalent salt molecules associated with a rod
segment of length $L=250\,r_0$.  The corresponding cell volume that contains
the mobile ions is then $V = L\pi R^2$ and the Debye length is $\ell_\rD={(8
  \pi \ell_\rB N/V)}^{-1/2}$.  Note that the line charge density
$\lambda=0.96\,e/r_0$ implies a number $M=240$ positive charges found on the
rod segment of length $L=250\,r_0$, and that the investigated $N$ ranges from
$0$ to $15\,000$.

\begin{figure}
  \vspace*{-0.5cm}
  \begin{center} 
\setlength{\unitlength}{0.1bp}
\begin{picture}(2339,1728)(0,0)
\special{psfile=fig4 llx=0 lly=0 urx=468 ury=403 rwi=4680}
\put(1344,50){\makebox(0,0){$r/r_0$}}
\put(50,964){%
\special{ps: gsave currentpoint currentpoint translate
270 rotate neg exch neg exch translate}%
\makebox(0,0)[b]{\shortstack{$P(r)$}}%
\special{ps: currentpoint grestore moveto}%
}
\put(2238,200){\makebox(0,0){50}}
\put(1808,200){\makebox(0,0){20}}
\put(1482,200){\makebox(0,0){10}}
\put(1156,200){\makebox(0,0){5}}
\put(725,200){\makebox(0,0){2}}
\put(400,200){\makebox(0,0){1}}
\put(300,1628){\makebox(0,0)[r]{1.0}}
\put(300,1362){\makebox(0,0)[r]{0.8}}
\put(300,1097){\makebox(0,0)[r]{0.6}}
\put(300,831){\makebox(0,0)[r]{0.4}}
\put(300,566){\makebox(0,0)[r]{0.2}}
\put(300,300){\makebox(0,0)[r]{0.0}}
\end{picture} \end{center}
  \caption{PB results (dotted curves) for the integrated charge distribution 
    function $P(r)$ for a system characterized by $r_0=\sigma$, $R/r_0 =
    61.9$, $\xi=2.1$ and $\lambda=0.96\,e/r_0$. Note that the number of rod
    counterions corresponding to rod length $L=250 r_0$ is $M=240$, while the
    number of 1:1 salt counterions (and co-ions) per length $L$ is, from
    bottom to top, $N=0, 104, 800, 3070$ and $15\,000$. The bold solid curve
    shows the locus of inflection points, i.e., the union of all inflection
    points of the functions $P(r)$.  The $\uparrow$-arrow marks the location
    of the salt-free Manning inflection point and the $\downarrow$-arrow shows
    where it joins one of the new salt inflection points. The branch of the
    locus between these two arrows indicates the range in which the concept of
    Manning condensation is meaningful. Observe that the functions are
    convex-up only within the grey-shaded region.}\label{fig:4}
\end{figure}
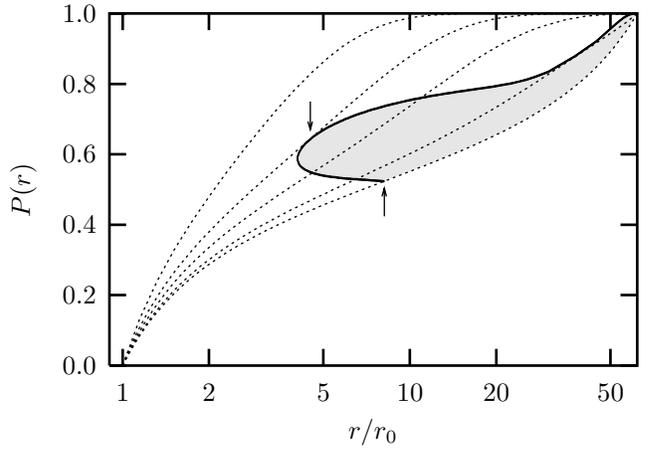

From the numerical solutions of $y(r)$ we have determined all inflection
points of $P(r)$ plotted against $\ln r$.  These inflection points are
solutions of the equation $\infd^2 P(r)/\infd (\ln r)^2 = 0$. In
figure~\ref{fig:4} we present the inflection points (bold solid curve)
starting from $N=0$ up to $N=3070$. For larger values of $N$ no further
inflection points are found. We also show the integrated charge distributions
(dotted) for $N=0$, $104$, $800$, $3070$ and $15\,000$, corresponding to Debye
lengths of $\ell_\rD/r_0=\infty$, $22.9$, $8.3$, $4.2$ and $1.9$,
respectively.

The location of the inflection point for $N=0$ coincides with $R_\rM$, thus
indicating a fraction of condensed counterions of $P(R_\rM)=1-1/\xi$.
Increasing $N$ by adding salt shifts the inflection point to smaller values of
$r$. That is, the layer of condensed counterions contracts, which is in accord
with other condensation criteria mentioned in section~\ref{sec_coucond}.
Importantly, the {\em amount} of condensed counterions is only marginally
increased in the presence of monovalent salt.  From a certain $N$ on (in
figure~\ref{fig:4} we find $N=104$) two more inflection points appear in
the high $r/r_0$ region. This happens typically for a corresponding Debye
length being of the order of the cell size itself, indicating the appearance
of a characteristic, salt induced, change in the convexity of $P$ (as a
function of $\ln r$).

Upon a further increase in $N$ one of the two new inflection points shifts
towards smaller $r/r_0$ values, finally fusing with the Manning inflection
point. Roughly speaking, we find the inflection points to vanish if the Debye
length characterizing the salt content becomes smaller than the radius of the
condensed layer. This suggests a breakdown of the need to distinguish between
condensed and uncondensed counterions once the typical salt screening length
interferes with the size of the condensed counterion layer.  Indeed, for a
very high salt content, where the Debye length is much smaller than the radius
of the rod, the solution of the PB equation would be the one of a charged
plane and one may consider all excess counterions being condensed no matter
what the charge density of the rod is.


\section{Comparison of PB theory with simulations: Added salt}\label{sec_compwith}

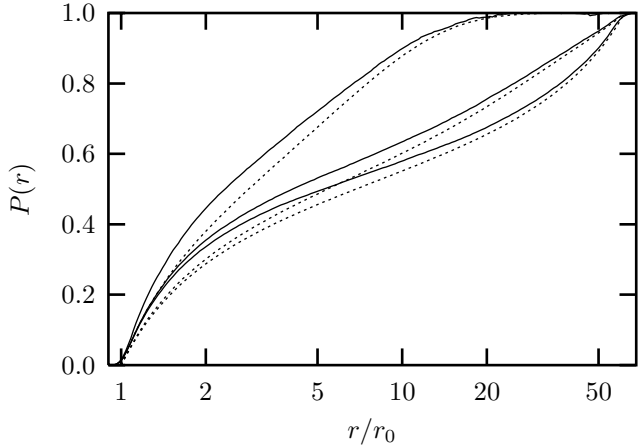
\begin{figure}
  \begin{center} 
\setlength{\unitlength}{0.1bp}
\begin{picture}(2339,1728)(0,0)
\special{psfile=fig5 llx=0 lly=0 urx=468 ury=403 rwi=4680}
\put(1344,50){\makebox(0,0){$r/r_0$}}
\put(50,964){%
\special{ps: gsave currentpoint currentpoint translate
270 rotate neg exch neg exch translate}%
\makebox(0,0)[b]{\shortstack{$P(r)$}}%
\special{ps: currentpoint grestore moveto}%
}
\put(2197,200){\makebox(0,0){50}}
\put(1776,200){\makebox(0,0){20}}
\put(1457,200){\makebox(0,0){10}}
\put(1138,200){\makebox(0,0){5}}
\put(717,200){\makebox(0,0){2}}
\put(398,200){\makebox(0,0){1}}
\put(300,1628){\makebox(0,0)[r]{1.0}}
\put(300,1362){\makebox(0,0)[r]{0.8}}
\put(300,1097){\makebox(0,0)[r]{0.6}}
\put(300,831){\makebox(0,0)[r]{0.4}}
\put(300,566){\makebox(0,0)[r]{0.2}}
\put(300,300){\makebox(0,0)[r]{0.0}}
\end{picture} \end{center}
  \caption{Simulated curves $P(r)$ for the systems with $N=0$, $104$ and $3070$ from 
    figure~\ref{fig:4} (solid lines) and the corresponding PB results (dotted
    curves).}\label{fig:5}
\end{figure}

\begin{figure}[t!]
  \vspace*{-0.35cm}
  \begin{center} 
\setlength{\unitlength}{0.1bp}
\begin{picture}(2339,1728)(0,0)
\special{psfile=fig6 llx=0 lly=0 urx=468 ury=403 rwi=4680}
\put(1344,50){\makebox(0,0){$r/r_0$}}
\put(50,964){%
\special{ps: gsave currentpoint currentpoint translate
270 rotate neg exch neg exch translate}%
\makebox(0,0)[b]{\shortstack{$P(r)$}}%
\special{ps: currentpoint grestore moveto}%
}
\put(2338,200){\makebox(0,0){17}}
\put(1979,200){\makebox(0,0){10}}
\put(1510,200){\makebox(0,0){5}}
\put(1359,200){\makebox(0,0){4}}
\put(1165,200){\makebox(0,0){3}}
\put(890,200){\makebox(0,0){2}}
\put(421,200){\makebox(0,0){1}}
\put(300,1628){\makebox(0,0)[r]{1.5}}
\put(300,1185){\makebox(0,0)[r]{1.0}}
\put(300,743){\makebox(0,0)[r]{0.5}}
\put(300,300){\makebox(0,0)[r]{0.0}}
\end{picture} \end{center}
  \caption{The integrated charge distribution function $P(r)$ 
    for a system characterized by $r_0=\sigma$, $R/r_0 = 15.5$, $\xi=4$,
    $\lambda/e=0.96/r_0$, $N=1000$ molecules of a divalent salt and $M=60$
    monovalent counterions corresponding to a rod segment of length $L=62.5 \:
    r_0$.  The simulation (solid curve) shows a pronounced
    overcharging-effect, in contrast to PB-theory (dotted
    curve).}\label{fig:6}
  \vspace*{-0.15cm}
  \begin{center} \vspace*{-1.5cm}\hspace*{0cm}
\setlength{\unitlength}{0.1bp}
\begin{picture}(3600,2160)(0,0)
\special{psfile=fig7 llx=0 lly=0 urx=720 ury=504 rwi=7200}
\put(1615,732){\makebox(0,0){$r/r_0$}}
\put(795,1022){%
\special{ps: gsave currentpoint currentpoint translate
270 rotate neg exch neg exch translate}%
\makebox(0,0)[b]{\shortstack{$y(r)$}}%
\special{ps: currentpoint grestore moveto}%
}
\put(1986,532){\makebox(0,0){5}}
\put(1871,532){\makebox(0,0){4}}
\put(1722,532){\makebox(0,0){3}}
\put(1512,532){\makebox(0,0){2}}
\put(1154,532){\makebox(0,0){1}}
\put(1020,1412){\makebox(0,0)[r]{0.5}}
\put(1020,1189){\makebox(0,0)[r]{0.0}}
\put(1020,966){\makebox(0,0)[r]{-0.5}}
\put(1020,743){\makebox(0,0)[r]{-1.0}}
\put(1369,50){\makebox(0,0){$r/r_0$}}
\put(50,964){%
\special{ps: gsave currentpoint currentpoint translate
270 rotate neg exch neg exch translate}%
\makebox(0,0)[b]{\shortstack{$n(r)\,r_0^3$}}%
\special{ps: currentpoint grestore moveto}%
}
\put(2338,200){\makebox(0,0){17}}
\put(1988,200){\makebox(0,0){10}}
\put(1531,200){\makebox(0,0){5}}
\put(1384,200){\makebox(0,0){4}}
\put(1194,200){\makebox(0,0){3}}
\put(927,200){\makebox(0,0){2}}
\put(469,200){\makebox(0,0){1}}
\put(350,1628){\makebox(0,0)[r]{0.03}}
\put(350,1185){\makebox(0,0)[r]{0.02}}
\put(350,743){\makebox(0,0)[r]{0.01}}
\put(350,300){\makebox(0,0)[r]{0.00}}
\end{picture} \end{center}
  \caption{The densities $n_{-2}(r)$ (solid line) and $n_{+2}(r)$ (dotted line)
    of negative and positive salt ions, respectively, for the same system
    presented in figure~\ref{fig:6}.  The inlay shows the electrostatic
    potential $y(r)$.}\label{fig:7}
\end{figure}

In this section we again compare the numerical results of the PB equation --
this time in the presence of salt -- with computer simulations. We
reinvestigate the systems in figure~\ref{fig:4} with $\xi = 2.1$, monovalent
counterions and number of salt molecules $N$ = 0, 104, and 3070 with respect
to a rod segment of length $L=250 r_0$. The results of the computer
simulations and the corresponding {\em mean-field} calculations are presented
in figure~\ref{fig:5}. Like in the salt-free case the computer simulations
show a more pronounced condensation effect towards the rod, which we again
attribute to ion-ion correlation effects.  Still, the shape of the
distribution functions remains qualitatively the same.  Note in particular
that the appearance and disappearance of two points of inflection at $N=104$
and $N=3070$ respectively, which leads to extremely small curvatures in the PB
distribution functions, also leads to very straight regions in the {\em
  measured} distribution functions. The crossover from Manning condensation to
screening, as described within PB theory, can thus be expected to be
essentially correct.

The PB approach fails to describe the physical situation if one or more of the
following points apply: {\em (i)} the electrostatic interactions are strong,
{\em (ii)} the counterions are multivalent and {\em (iii)} the density is
high. A simulation under such conditions can be inspected in
figure~\ref{fig:6}.  In this system we have $r_0=\sigma$, $R/r_0 \approx
15.5$, $\lambda=0.96e/r_0$, $\xi=4$, 60 monovalent counterions, and 1000
molecules of a 2:2 salt (giving a Debye length of roughly $0.33\,r_0$, i.e.,
smaller than the ion diameter). Here $P(r)$ overshoots unity, showing a charge
reversal of the rod at distances around $r \approx 1.5\,r_0$, while the simple
PB prediction is clearly {\em qualitatively} off.  This phenomenon is usually
referred to as {\em overcharging} and has been predicted for the primitive
cell model first from hypernetted chain calculations\cite{tovar85a} and later
by a modified Poisson-Boltzmann approach.\cite{Das95Das97}

Since $P(R)=1$, the overshooting above 1 at small distances implies the
existence of a range of $r$-values at which the mobile ion system is locally
{\em positively} charged (i.e., with the same charge as the rod), such that
$P(r)$ can eventually decay to 1. This is seen in figure~\ref{fig:6}, which
shows that $n_{+2}(r)>n_{-2}(r)$ at $r \approx 2\,r_0$. Since $P(1.5)\approx
1.45$, the rod and its innermost layer of condensed ions could be viewed as an
effective rod of radius $1.5\,r_0$ which is negatively charged with Manning
parameter $\xi=1.8$. Since this value is again larger than 1, it entails ion
condensation (this time of the positive ions). In fact, it even leads to a
second overcharging, as can clearly be seen in figure~\ref{fig:6}, where
$P(r)$ -- in decaying from 1.45 -- overshoots the value of 1 again.
Overcharging can thus give rise to layering; in the presented example no less
than three layers can clearly be made out. These local charge oscillations
also reflect themselves in oscillations of the electrostatic potential, as
demonstrated in the inset of figure~\ref{fig:7}. Note that these
oscillating potentials will also have pronounced effects on the interaction
{\em between} such rigid polyelectrolytes.


\section*{Conclusions}

We have revisited counterion condensation with and without added salt for a
solution of rigid polyelectrolytes within the cell model approximation.  It
was confirmed that on the level of PB theory with no added salt a simple
geometric method locates the condensation radius $R_\rM$ as well as the
fraction of condensed counterions.  This geometric method consists in finding
the inflection point of the integrated probability distribution $P$ plotted as
a function of $\ln r$.  Without added salt, the locations of the inflection
point and the Manning radius $R_\rM$, where a fraction of $1-1/\xi$
counterions are condensed, are identical.

A key point in the present work was to extend the inflection point based
counterion condensation criterion to the case of added salt and to compare its
implications as predicted by PB theory and by computer simulations.  Our
motivation for introducing this new condensation criterion was {\em (i)} to
avoid fixing by definition the amount of condensed counterions, {\em (ii)} to
reproduce the salt-free PB limit, namely $P(R_\rM)=1-1/\xi$, and {\em (iii)}
to predict a counterion condensation breakdown in the high salt limit. We are
not aware of any other condensation criterion that fulfills all these
requirements at the same time.

Upon addition of monovalent salt we found PB theory to predict counterion
condensation within a somewhat smaller region around the charged rod. This is
in accord with other studies\cite{mills85,murthy85} using different
condensation criteria.\cite{Gueron80,Lamm94,Jayaram94}  Importantly, the
fraction of condensed counterions did not exhibit a strong salt dependence
thus supporting a principal idea in Manning condensation: The number of
condensed counterions does not depend on salt. Yet, this observation must
breakdown in the high salt limit, where the screening length of the salt
becomes of the order of the size of the condensed counterion layer. Indeed,
our new condensation criterion naturally predicts this to happen.

All these behaviors were well reproduced by our computer simulations.  In
fact, the agreement between simulation and the corresponding {\em mean-field}
level calculation is remarkably good for systems with a Manning parameter
$\xi<1$. Upon increasing the counterion density, valence, or the Manning
parameter, the simulations predict consistently a somewhat stronger
condensation.  We have argued that this finding is due to ionic correlations
not present in PB theory. To test this assumption, we have also performed a
simulation of a system in a highly concentrated divalent salt environment.
Here we clearly saw the phenomenon of overcharging, which a corresponding PB
calculation was unable to reproduce.

We note finally that our MD simulations suggest the usability of the
counterion condensation criterion beyond the PB cell model approximation.  In
fact, we have observed inflection points in the integrated probability
distribution functions in systems very distinct from the rigid rods employed
in the present study, namely for flexible polyelectrolytes in the presence of
multivalent counterions and for flexible polyelectrolytes in poor solvents
with monovalent counterions.\cite{deserno99c}


\section*{Acknowledgments}
CH and MD thank G.\ Manning for useful conversations and for bringing
Belloni's work to our attention.  This work has been started at the ITP, Santa
Barbara, whose financial support under NFS grant No. PHY-94-07194 is
gratefully acknowledged. SM wishes to thank SFB 197 for its support.


\section*{Appendix: Details of the simulation}\label{appendix}

The system used to study counterion condensation consists of a cubic
simulation box of length $L_\rb$, a charged rod along the main diagonal, the
appropriate amount of coun\-ter\-ions necessary for electric neutrality and
possibly some additional salt. Upon switching on 3D-periodic boundary
conditions a triangular array of infinitely long charged rods is modelled.
Note that in PB theory we approximate the corresponding Wigner-Seitz-cell by a
cylindrically symmetric unit cell of the same volume (implying
$R=L_\rb/\sqrt{\pi \surd 3}$) thus rendering the PB equation one-dimensional.

Apart from electrostatic interactions all ions are subject to a purely
repulsive Lennard-Jones potential, giving an excluded volume and a
corresponding ion-diameter $\sigma$. Further, the rod is modelled as an
immobile string of such spheres, having the separation $1.042\sigma$. The
distance of closest approach to the cylinder, i.e.\ $r_0$, turns out to be
essentially $1 \: \sigma$.

The electrostatic interactions in this periodic boundary geometry were
computed with the help of P$^3$M routines\cite{P3M} and a Langevin
thermostat\cite{LangDyn} combined with a
velocity-Verlet-integrator\cite{AlTi97} (with timestep 0.0125 in LJ units) was
implemented to drive the system into the canonical state. The number of MD
steps varied between $8\times10^5$ (for the systems from figure
~\ref{fig:2}) up to $6.4\times10^6$ (for the system from figures
~\ref{fig:6} and \ref{fig:7}) and the saturation of the electrostatic
energy was used to test for equilibration. A more detailed description of our
simulation method will be presented in a forthcoming
publication.\cite{deserno99b}



\end{document}